\documentclass[english,pra,twocolumn,showpacs]{revtex4}
\usepackage[T1]{fontenc}
\usepackage[latin9]{inputenc}
\usepackage{amsmath}
\usepackage{amssymb}
\usepackage{graphicx}
\usepackage{esint}

\makeatletter
\@ifundefined{textcolor}{}
{%
 \definecolor{BLACK}{gray}{0}
 \definecolor{WHITE}{gray}{1}
 \definecolor{RED}{rgb}{1,0,0}
 \definecolor{GREEN}{rgb}{0,1,0}
 \definecolor{BLUE}{rgb}{0,0,1}
 \definecolor{CYAN}{cmyk}{1,0,0,0}
 \definecolor{MAGENTA}{cmyk}{0,1,0,0}
 \definecolor{YELLOW}{cmyk}{0,0,1,0}
}

\usepackage[dvipdfm,
            colorlinks,
            linkcolor=red,
            citecolor=blue
            ]{hyperref}

\makeatother

\usepackage{babel}
\begin{document}

\title{Two-body state with $p$-wave interaction in one-dimensional waveguides
under transversely anisotropic confinement }

\author{Tian-You Gao$^{1}$}

\author{Shi-Guo Peng$^{1}$}

\email{pengshiguo@gmail.com}

\author{Kaijun Jiang$^{1,2}$}

\email{kjjiang@wipm.ac.cn}

\affiliation{$^{1}$State Key Laboratory of Magnetic Resonance and Atomic and
Molecular Physics, Wuhan Institute of Physics and Mathematics, Chinese
Academy of Sciences, Wuhan 430071, China }

\affiliation{$^{2}$Center for Cold Atom Physics, Chinese Academy of Sciences,
Wuhan 430071, China}

\date{\today}
\begin{abstract}
We theoretically study two atoms with $p$-wave interaction in a one-dimensional
waveguide, and investigate how the transverse anisotropy of the confinement
affects the two-body state, especially, the properties of the resonance.
For bound-state solution, we find there are totally three two-body
bound states due to the richness of the orbital magnetic quantum number
of $p$-wave interaction, while only one bound state is supported
by $s$-wave interaction. Two of them become nondegenerate due to
the breaking of the rotation symmetry under transversely anisotropic
confinement. For scattering solution, the effective one-dimensional
scattering amplitude and scattering length are derived. We find the
position of the $p$-wave confinement-induced resonance shifts apparently
as the transverse anisotropy increases. In addition, a two-channel
mechanism for confinement-induced resonance in a one-dimensional waveguide
is generalized to $p$-wave interaction, which was proposed only for
$s$-wave interaction before. All our calculations are based on the
parameterization of the $^{40}$K atom experiments, and can be confirmed
in future experiments.
\end{abstract}


\maketitle

\section{Introduction}

Nowadays, the confinement-induced resonance (CIR) as one of the most
intriguing phenomena in low-dimensional systems has attracted a great
deal of interest. It was first predicted by Olshanii when considering
the two-body $s$-wave scattering problem in quasi-one-dimensional
(quasi-1D) waveguides \cite{Olshanii1998}. Subsequently, this study
was extended to quasi-two-dimensional (quasi-2D) systems \cite{Petrov2000B,Petrov2001I}.
In the past decade, an impressive amount of experimental and theoretical
efforts have been devoted to confirm the existence of CIRs and explore
the important consequences \cite{Bergeman2003A,Granger2004T,Moritz2005C,Gunter2005P,Pricoupenko2008R,Haller2009R,Haller2010C,Peng2010C,Zhang2011C,Frohlich2011R,Peng2011C,Sala2012I,Giannakeas2011R,Peng2012T,Peng2014M}.
To date, CIR has already become a fundamental technique in studying
strongly interacting low-dimensional quantum gases.

The $p$-wave interaction is of particular interest, since it is the
simplest high-partial-wave interaction with non-zero orbital angular
momentum $l=1$. This leads to a spatial anisotropic scattering in
few-body physics, and open a new method to manipulate resonant cold
atomic interactions by using the magnetic field vector, as indicated
in our previous work \cite{Peng2014M}. For many-body physics, the
high-partial-wave interaction may also result in physically abundant
quantum phase transitions, which are absent in $s$-wave interaction
\cite{Ohashi2005B,Ho2005F,Cheng2005A,Iskin2006E}. In addition, the
$p$-wave scattering dominates the low-energy interaction between
spin-polarized Fermi atoms (all in one hyperfine state) due to the
Fermi-Dirac statistics, which provide an ideal candidate to study
the $p$-wave scattering properties in cold atoms \cite{Zhang2004P,Ticknor2004M}.

For low-dimensional quantum systems, the external trap potential is
one another freedom to manipulate two-body resonant interactions,
and it is interesting to identify how CIR is affected when the external
confinement changes. For example, in 1D Bose $^{133}$Cs atoms with
$s$-wave interaction, it is found that the transverse harmonic anisotropy
shifts the position of CIR \cite{Peng2010C,Zhang2011C}, besides,
CIR splits due to the anharmonicity of the transverse trap \cite{Haller2010C,Peng2011C,Sala2012I}.
Naturally, it is also important to investigate how the external confinement
affects the low-dimensional resonant $p$-wave interaction between
spin-polarized Fermi atoms, such as $^{40}$K atoms. 

In this paper, we study the influence of the transverse anisotropy
on two-body state with $p$-wave interaction, especially, the scattering
resonance in quasi-1D waveguides. We solve the two-body problem in
a quasi-1D waveguide under transversely anisotropic confinement. The
$p$-wave interaction is modeled by the contact pseudopotential including
the effective range \cite{Peng2011H}. This model neglects the spatial
anisotropy of $p$-wave interaction , which is not the focus of this
work. However, it is already sufficient enough to study how the external
trap affects the $p$-wave CIR. Due to the non-zero orbital angular
momentum, we find there are totally three bound states for $p$-wave
interaction in the waveguide, while the $s$-wave pseudopotential
supports only one two-body bound state. Two of them become nondegenerate
as the transverse confinement changes from isotropic to anisotropic,
which breaks the rotation symmetry around the waveguide. For scattering
solution, we calculate the effective 1D scattering amplitude as well
as the 1D scattering length, and predict the scattering resonance
for any transverse anisotropy, whose position shows an apparent shift
with increasing anisotropy. In addition, we present an effective two-channel
mechanism for $p$-wave CIR, which was first proposed only for $s$-wave
interaction \cite{Bergeman2003A}. In this two-channel picture, the
transverse ground state and the remaining transverse excited modes
play roles of open and closed channels, respectively. When a bound
state in the closed channel exists, and becomes energetically degenerate
with the scattering threshold of the open channel, a scattering resonance
is expected. All our calculations are based on the parameterization
of the $^{40}$K atom experiments \cite{Ticknor2004M,Gunter2005P},
and can be confirmed in future experiments.

In the following, we first study the properties of two-body bound
states with $p$-wave interaction in a transversely anisotropic waveguide
(Sec. \ref{sec:BoundStateSolutions}), and then present the scattering
solution in Sec. \ref{sec:ScatteringSolutions} as well. We calculate
the effective 1D scattering amplitude and 1D scattering length, and
discuss how the transverse anisotropy of the confinement affects the
$p$-wave CIR in details. In Sec. \ref{sub:TheEffectiveTwoChannelMechanism},
a two-channel mechanism is presented for $p$-wave CIR. The main results
are concluded in Sec. \ref{sec:Conclusions}.

\section{bound-state solutions \label{sec:BoundStateSolutions}}

In order to simplify the problem, let us consider two atoms confined
in a quasi-1D waveguide with a tight transverse harmonic potential
(in $x-y$ plane). The transverse trap frequencies are $\omega_{x,y}$
, and the atoms can freely move along $z$ axis. The transverse anisotropy
is characterized by $\eta=\omega_{x}/\omega_{y}$ . Unlike the $s$-wave
interaction, the $p$-wave interaction is strongly energy-dependence
due to the narrow-width property \cite{Ticknor2004M}. Thus, one additional
parameter named effective range should be included \cite{Peng2011H,Yip2008E,Suzuki2009T,Idziaszek2009A}.
In harmonic traps, the center-of-mass (c.m.) motion is decoupled from
the relative part, and we only need to solve the relative motion,
while the c.m. motion is a simple harmonic oscillator. By dropping
the c.m. motion off, the relative motion of two atoms is described
by the following Hamiltonian,
\begin{equation}
\mathcal{H}=-\frac{\hbar^{2}}{2\mu}\frac{\partial^{2}}{\partial z^{2}}+\mathcal{H}_{\perp}+\mathcal{V}_{1},\label{eq:2.1}
\end{equation}
where
\begin{equation}
\mathcal{H}_{\perp}=-\frac{\hbar^{2}}{2\mu}\left(\frac{\partial^{2}}{\partial x^{2}}+\frac{\partial^{2}}{\partial y^{2}}\right)+\frac{1}{2}\mu\omega^{2}\left(\eta^{2}x^{2}+y^{2}\right),\label{eq:2.2}
\end{equation}
 $\mu$ is the reduced mass, and $\omega$ is the trap frequency in
$y$ aixs (we omit the subscript $y$ without ambiguity). The interatomic
interaction $\mathcal{V}_{1}\left(\mathbf{r}\right)$ is modeled by
the $p$-wave pseudopotential \cite{Peng2011H}, 
\begin{equation}
\mathcal{V}_{1}\left(\mathbf{r}\right)=\frac{\pi\hbar^{2}}{\mu}\left(\frac{1}{v_{1}}-\frac{1}{2}r_{1}k^{2}\right)^{-1}\overleftarrow{\nabla}\delta\left(\mathbf{r}\right)\frac{\partial^{3}}{\partial r^{3}}r^{3}\overrightarrow{\nabla},\label{eq:2.3}
\end{equation}
where $v_{1}$ and $r_{1}$ are the three-dimensional (3D) $p$-wave
scattering volume and effective range (with a dimension of inverse
length), respectively, which can be tuned by using $p$-wave Feshbach
resonances \cite{Ticknor2004M,Zhang2004P}. $k$ is the relative wavenumber,
related to the relative energy of two atoms as $E=\hbar^{2}k^{2}/\left(2\mu\right)$
. The symbol $\overleftarrow{\nabla}$ ($\overrightarrow{\nabla}$)
denotes the gradient operator that acts to the left (right) of the
pseudopotential.

For the bound-state problem, the wavefunction can formally be written
as
\begin{equation}
\psi\left(\mathbf{r}\right)=-\int d^{3}\mathbf{r}^{\prime}G_{E}\left(\mathbf{r},\mathbf{r}^{\prime}\right)\mathcal{V}_{1}\left(\mathbf{r}^{\prime}\right)\psi\left(\mathbf{r}^{\prime}\right),\label{eq:2.4}
\end{equation}
 where $G_{E}\left(\mathbf{r},\mathbf{r}^{\prime}\right)$ is the
single-particle Green's function with energy $E$ , and satisfies
\begin{equation}
\left(-\frac{\hbar^{2}}{2\mu}\frac{\partial^{2}}{\partial z^{2}}+\mathcal{H}_{\perp}-E\right)G_{E}\left(\mathbf{r},\mathbf{r}^{\prime}\right)=\delta\left(\mathbf{r}-\mathbf{r}^{\prime}\right).\label{eq:2.5}
\end{equation}
The Green's function can be expanded in series of the eigen-states
of non-interacting Hamiltonian as,
\begin{multline}
G_{E}\left(\mathbf{r},\mathbf{r}^{\prime}\right)=\sum_{n_{1}=0}^{\infty}\sum_{n_{2}=0}^{\infty}\int_{-\infty}^{\infty}dk_{z}\\
C_{n_{1}n_{2}}\left(k_{z}\right)\phi_{n_{1}}\left(\frac{\sqrt{\eta}x}{d}\right)\phi_{n_{2}}\left(\frac{y}{d}\right)\frac{e^{ik_{z}z}}{\sqrt{2\pi}},\label{eq:2.6}
\end{multline}
where $\phi_{\nu}\left(\cdot\right)$ is the eigen-state of 1D harmonic
oscillator, $d=\sqrt{\hbar/\mu\omega}$ is the harmonic length in
$y$ aixs, and $k_{z}$ is the wavenumber along the waveguide ($z$
axis). Inserting Eq.(\ref{eq:2.6}) into Eq.(\ref{eq:2.5}) and using
the completeness of the eigenstates of the non-interacting Hamiltonian,
we may easily obtain the coefficients $C_{n_{1}n_{2}}\left(k_{z}\right)$
, and then after some straightforward algebra (the derivation is similar
to Eq.(24) in \cite{Peng2010C}), the Green's function takes the following
integral representation,
\begin{multline}
G_{E}\left(\mathbf{r},\mathbf{r}^{\prime}\right)=\frac{1}{\pi^{3/2}d^{3}\hbar\omega}\int_{0}^{\infty}\frac{d\tau\, e^{\epsilon\tau}}{\sqrt{2\tau}}\exp\left[-\frac{\left(z-z^{\prime}\right)^{2}}{2d^{2}\tau}\right]\\
\times\frac{\sqrt{\eta}}{\sqrt{1-e^{-2\eta\tau}}}\exp\left[\eta\frac{2x^{\prime}x-\left(x^{\prime2}+x^{2}\right)\cosh\left(\eta\tau\right)}{2d^{2}\sinh\left(\eta\tau\right)}\right]\\
\times\frac{1}{\sqrt{1-e^{-2\tau}}}\exp\left[\frac{2y^{\prime}y-\left(y^{\prime2}+y^{2}\right)\cosh\tau}{2d^{2}\sinh\tau}\right],\label{eq:2.7}
\end{multline}
which is valid for $\epsilon\equiv E/\hbar\omega-\left(\eta+1\right)/2<0$
. Combining Eqs.(\ref{eq:2.3}) and (\ref{eq:2.4}), the bound-state
wavefunction takes the form
\begin{multline}
\psi\left(\mathbf{r}\right)=-\frac{\pi\hbar^{2}}{\mu}\left(\frac{1}{v_{1}}-\frac{1}{2}r_{1}k^{2}\right)^{-1}\\
\cdot\sum_{n}\mathcal{R}_{n}\left[\frac{\partial}{\partial r_{n}^{\prime}}G_{E}\left(\mathbf{r},\mathbf{r}^{\prime}\right)\right]_{r^{\prime}=0},\label{eq:2.8}
\end{multline}
where the summation is over $n=x,y,z$ , and $r_{x}\equiv x,\, r_{y}\equiv y,\, r_{z}\equiv z$
. Here, we have defined the coefficent
\begin{equation}
\mathcal{R}_{n}\equiv\left[\frac{\partial^{3}}{\partial r^{3}}r^{3}\frac{\partial}{\partial r_{n}}\psi\left(\mathbf{r}\right)\right]_{r=0}.\label{eq:2.9}
\end{equation}
Acting $\frac{\partial^{3}}{\partial r^{3}}r^{3}\frac{\partial}{\partial r_{n}}$
on both sides of Eq.(\ref{eq:2.8}) and setting $r=0$ , we obtain
the following secular equation,
\begin{multline}
\sum_{n}\left[-\frac{\pi\hbar^{2}}{\mu}\left(\frac{1}{v_{1}}-\frac{1}{2}r_{1}k^{2}\right)^{-1}\mathcal{A}_{ln}-\delta_{ln}\right]\mathcal{R}_{n}=0,\\
\;\left(l=x,y,z\right),\label{eq:2.10}
\end{multline}
 where 
\begin{equation}
\mathcal{A}_{ln}\equiv\left[\frac{\partial^{3}}{\partial r^{3}}r^{3}\frac{\partial^{2}}{\partial r_{l}\partial r_{n}^{\prime}}G_{E}\left(\mathbf{r},\mathbf{r}^{\prime}\right)\right]_{r=r^{\prime}=0}.\label{eq:2.11}
\end{equation}
 For any nonzero vector $\mathbf{r}$ , it is easy to show that
\begin{multline}
\left[\frac{\partial G_{E}\left(\mathbf{r},\mathbf{r}^{\prime}\right)}{\partial r_{n}^{\prime}}\right]_{r^{\prime}=0}=\frac{r_{n}}{\pi^{3/2}d^{5}\hbar\omega}\left[\mathcal{F}_{n}\left(\epsilon,\mathbf{r}\right)\right.\\
\left.+\frac{\sqrt{\pi}}{2}\left(\frac{d^{3}}{r^{3}}+\frac{E}{\hbar\omega}\cdot\frac{d}{r}\right)\right],\label{eq:2.12}
\end{multline}
 in which\begin{widetext}
\begin{eqnarray}
\mathcal{F}_{x}\left(\epsilon,\mathbf{r}\right) & = & \int_{0}^{\infty}d\tau\left\{ \frac{\eta^{3/2}e^{\epsilon\tau}}{\sqrt{2}\sinh\left(\eta\tau\right)}\frac{\exp\left[-\frac{\eta x^{2}}{2d^{2}\tanh\left(\eta\tau\right)}-\frac{y^{2}}{2d^{2}\tanh\tau}-\frac{z^{2}}{2d^{2}\tau}\right]}{\sqrt{\tau\left(1-e^{-2\eta\tau}\right)\left(1-e^{-2\tau}\right)}}-\frac{1}{2\sqrt{2}}\left(\frac{1}{\tau^{5/2}}+\frac{E}{\hbar\omega}\frac{1}{\tau^{3/2}}\right)e^{-\frac{r^{2}}{2d^{2}\tau}}\right\} ,\label{eq:2.13}\\
\mathcal{F}_{y}\left(\epsilon,\mathbf{r}\right) & = & \int_{0}^{\infty}d\tau\left\{ \frac{\eta^{1/2}e^{\epsilon\tau}}{\sqrt{2}\sinh\tau}\frac{\exp\left[-\frac{\eta x^{2}}{2d^{2}\tanh\left(\eta\tau\right)}-\frac{y^{2}}{2d^{2}\tanh\tau}-\frac{z^{2}}{2d^{2}\tau}\right]}{\sqrt{\tau\left(1-e^{-2\eta\tau}\right)\left(1-e^{-2\tau}\right)}}-\frac{1}{2\sqrt{2}}\left(\frac{1}{\tau^{5/2}}+\frac{E}{\hbar\omega}\frac{1}{\tau^{3/2}}\right)e^{-\frac{r^{2}}{2d^{2}\tau}}\right\} ,\label{eq:2.14}\\
\mathcal{F}_{z}\left(\epsilon,\mathbf{r}\right) & = & \int_{0}^{\infty}d\tau\left\{ \frac{\eta^{1/2}e^{\epsilon\tau}}{\sqrt{2}\tau}\frac{\exp\left[-\frac{\eta x^{2}}{2d^{2}\tanh\left(\eta\tau\right)}-\frac{y^{2}}{2d^{2}\tanh\tau}-\frac{z^{2}}{2d^{2}\tau}\right]}{\sqrt{\tau\left(1-e^{-2\eta\tau}\right)\left(1-e^{-2\tau}\right)}}-\frac{1}{2\sqrt{2}}\left(\frac{1}{\tau^{5/2}}+\frac{E}{\hbar\omega}\frac{1}{\tau^{3/2}}\right)e^{-\frac{r^{2}}{2d^{2}\tau}}\right\} .\label{eq:2.15}
\end{eqnarray}
\end{widetext} Substituting Eq.(\ref{eq:2.12}) into Eq.(\ref{eq:2.11}),
it directly yields
\begin{equation}
\mathcal{A}_{ln}=\frac{6}{\pi^{3/2}d^{5}\hbar\omega}\mathcal{F}_{n}\left(\epsilon,0\right)\delta_{ln},\label{eq:2.16}
\end{equation}
 where $\delta_{ln}$ is Kronecker delta function. Combining Eqs.(\ref{eq:2.10})
and (\ref{eq:2.16}), the vanish of the determinant of the coefficient
matrix in Eq.(\ref{eq:2.10}) determines the binding energy, which
yields 
\begin{equation}
\frac{d^{3}}{v_{1}}=-\frac{6}{\sqrt{\pi}}\mathcal{F}_{l}\left(\epsilon,0\right)+dr_{1}\cdot\frac{E}{\hbar\omega},\,\left(l=x,y,z\right).\label{eq:2.17}
\end{equation}
 Obviously, there are totally three bound states for $p$-wave interaction,
while the $s$-wave pseudopotential supports only one two-body bound
state. This is resulted from the richness of the orbital magnetic
quantum number of the $p$-wave interaction. The corresponding (un-normalized)
bound-state wavefunction is
\begin{equation}
\psi_{l}\left(\mathbf{r}\right)=r_{l}\left[\frac{1}{r^{3}}+\frac{E}{\hbar\omega}\cdot\frac{1}{rd^{2}}+\frac{2}{\sqrt{\pi}d^{3}}\mathcal{F}_{l}\left(\epsilon,\mathbf{r}\right)\right],\;\left(l=x,y,z\right).\label{eq:2.18}
\end{equation}
 Obviously, these bound states can be classified as two kinds by the
transverse parity (in $x$ or $y$ axis), i.e., $\psi_{x,y}\left(\mathbf{r}\right)$
with odd transverse parity and $\psi_{z}\left(\mathbf{r}\right)$
with even transverse parity. For a transversely isotropic confinement
($\eta=1$), there is a rotation symmetry around the $z$ axis, and
$\psi_{x}\left(\mathbf{r}\right)$ and $\psi_{y}\left(\mathbf{r}\right)$
are degenerate. Thus, at small separation, i.e., $\mathbf{r}\approx0$
, we can construct another set of eigen-wavefunctions with specific
orbital magnetic quantum number as superposition of $\psi_{x}\left(\mathbf{r}\right)$
, $\psi_{y}\left(\mathbf{r}\right)$ , and $\psi_{z}\left(\mathbf{r}\right)$
, i.e.,

\begin{eqnarray}
\psi_{m=\pm1}\left(\mathbf{r}\right) & = & \mp\sqrt{\frac{3}{8\pi}}\left[\psi_{x}\left(\mathbf{r}\right)\pm i\psi_{y}\left(\mathbf{r}\right)\right],\label{eq:2.19}\\
\psi_{m=0}\left(\mathbf{r}\right) & = & \sqrt{\frac{3}{4\pi}}\psi_{z}\left(\mathbf{r}\right),\label{eq:2.20}
\end{eqnarray}
or explicitly,
\begin{equation}
\psi_{m=\pm1}\left(\mathbf{r}\right)=\left[\frac{1}{r^{2}}+\frac{E}{d^{2}\hbar\omega}+\frac{2r}{\sqrt{\pi}d^{3}}\mathcal{F}_{x}\left(\epsilon,0\right)\right]Y_{1\pm1}\left(\hat{\mathbf{r}}\right),\label{eq:2.21}
\end{equation}
\begin{equation}
\psi_{m=0}\left(\mathbf{r}\right)=\left[\frac{1}{r^{2}}+\frac{E}{d^{2}\hbar\omega}+\frac{2r}{\sqrt{\pi}d^{3}}\mathcal{F}_{z}\left(\epsilon,0\right)\right]Y_{10}\left(\hat{\mathbf{r}}\right),\label{eq:2.22}
\end{equation}
 with specific orbital magnetic quantum number $m=0,\pm1$ , and
\begin{eqnarray}
Y_{1\pm1}\left(\hat{\mathbf{r}}\right) & = & \mp\sqrt{\frac{3}{8\pi}}\frac{x\pm iy}{r},\label{eq:2.23}\\
Y_{10}\left(\hat{\mathbf{r}}\right) & = & \sqrt{\frac{3}{4\pi}}\frac{z}{r}\label{eq:2.24}
\end{eqnarray}
 are spherical harmonic functions. As the transverse confinement becomes
anisotropic, the rotation symmetry around the $z$ axis is broken.
This results in the nondegeneracy of $\psi_{x}\left(\mathbf{r}\right)$
and $\psi_{y}\left(\mathbf{r}\right)$ , and they are no longer the
superposition of those with specific orbital magnetic quantum number
at small $\mathbf{r}$ . 

\begin{figure}
\includegraphics[width=1\columnwidth]{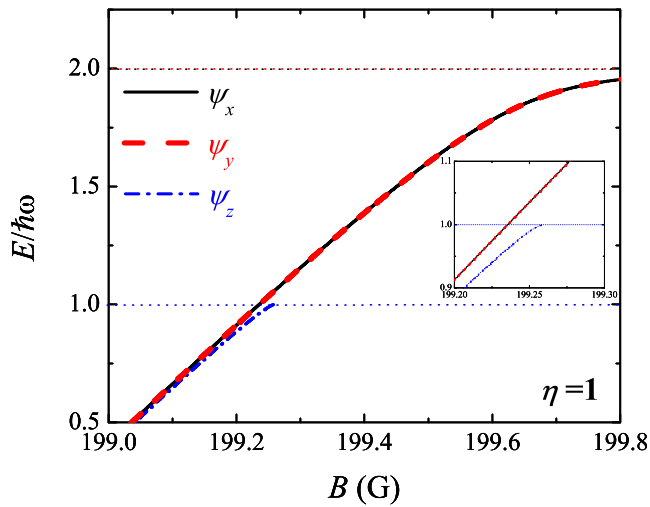}

\caption{(Color online) The bound-state energy of two $^{40}$K atoms in the
$\left|F=\frac{9}{2},m_{F}=-\frac{7}{2}\right\rangle $ hyperfine
state confined in a transversely isotropic\emph{ }waveguide near the
$p$-wave Feshbach resonance centered at $B_{0}=198.8$G in three
dimension \cite{Ticknor2004M,Gunter2005P}. The dotted horizontal
lines are the thresholds that the bound states merge into the continuum
as the magnetic field increases. The magnetic field $B$ dependence
of the scattering volume $v_{1}$ as well as the effective range $r_{1}$
for $^{40}$K is from Ref.\cite{Ticknor2004M}. The insert shows the
details of the curves in a smaller magnetic field range for eye clarity.}

\label{BoundState(eta=00003D1)}
\end{figure}

\begin{figure}
\includegraphics[width=1\columnwidth]{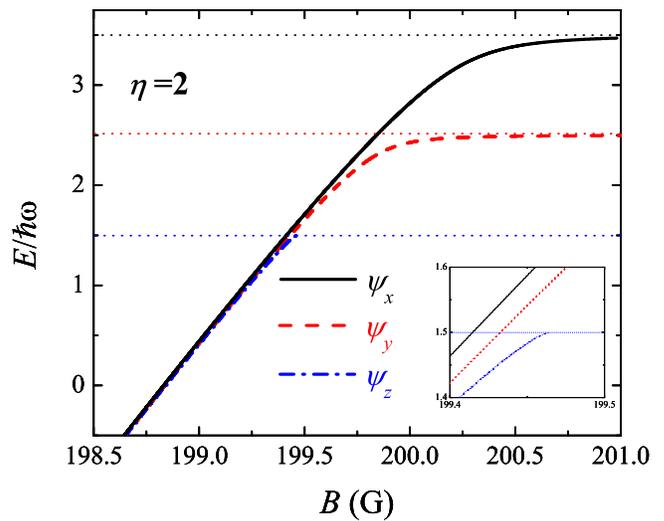}

\caption{(Color online) The same as Fig.\ref{BoundState(eta=00003D1)} but
for a transversely anisotropic confinement, i.e., $\eta=2$ .}

\label{BoundState(eta=00003D2)}
\end{figure}

We predict the binding energy of two $^{40}$K atoms in the $\left|F=\frac{9}{2},m_{F}=-\frac{7}{2}\right\rangle $
hyperfine state as functions of the magnetic field near the $p$-wave
Feshbach resonance centered at $B_{0}=198.8$G in three dimension
(3D)\cite{Ticknor2004M,Gunter2005P}, confined in a transversely isotropic
waveguide (Fig.\ref{BoundState(eta=00003D1)}) as well as in a transversely
anisotropic waveguide (Fig.\ref{BoundState(eta=00003D2)}). For the
isotropic confinement ($\eta=1$), $\psi_{x}\left(\mathbf{r}\right)$
and $\psi_{y}\left(\mathbf{r}\right)$ are degenerate as we anticipate
(see Fig.\ref{BoundState(eta=00003D1)}), and they become distinguishable
in energy due to the rotation symmetry breaking around the $z$ axis
when the transverse trap becomes anisotropic (see Fig.\ref{BoundState(eta=00003D2)}).
In addition, as the magnetic field increases, these bound states gradually
merge into the continuum at different energy thresholds, i.e., $\left(3\eta+1\right)/2$
, $\left(\eta+3\right)/2$ , $\left(\eta+1\right)/2$ , respectively.

\section{scattering solutions and confinement-induced resoannces\label{sec:ScatteringSolutions}}

In this section, let us consider the low-energy $p$-wave scattering
problem in the 1D waveguide with energy just above the transverse
zero-point energy, i.e., $\epsilon\rightarrow0^{+}$ . According to
the Lippman-Schwinger equation, the scattering solution of the Hamiltonian
(\ref{eq:2.1}) can formally be written as

\begin{equation}
\psi(\mathbf{r})=\psi_{0}\left(\mathbf{r}\right)-\int d^{3}\mathbf{r}^{\prime}G_{E}\left(\mathbf{r},\mathbf{r}^{\prime}\right)\mathcal{V}_{1}\left(\mathbf{r}^{\prime}\right)\psi\left(\mathbf{r}^{\prime}\right),\label{eq:3.1}
\end{equation}
where $\psi_{0}\left(\mathbf{r}\right)$ is the incident wavefunction.
Since the energy of the relative motion is just above the transverse
zero-point energy, the atoms should enter from the transverse ground
state, and the incident wavefunction takes the form of 
\begin{equation}
\psi_{0}\left(\mathbf{r}\right)=\phi_{0}\left(\frac{\sqrt{\eta}x}{d}\right)\phi_{0}\left(\frac{y}{d}\right)i\sin\left(k_{z}z\right).\label{eq:3.2}
\end{equation}
Here, we have considered the exchanging anti-symmetry of two identical
fermions. Remind that $\phi_{0}\left(\cdot\right)$ is the 1D harmonic
ground-state wavefunction and $k_{z}=\sqrt{2\epsilon}/d$ . The two-body
Green's function $G_{E}\left(\mathbf{r},\mathbf{r}^{\prime}\right)$
is the simple analytical continuation of Eq.(\ref{eq:2.7}) from $\epsilon<0$
to $\epsilon\approx0^{+}$ . Substituting the $p$-wave pseudopotential
(\ref{eq:2.3}) into Eq.(\ref{eq:3.1}), and at large separation,
i.e., $r\rightarrow\infty$ , we find the scattering wavefunction
(\ref{eq:3.1}) behaves as
\begin{multline}
\psi(\mathbf{r})\approx\phi_{0}\left(\frac{\sqrt{\eta}x}{d}\right)\phi_{0}\left(\frac{y}{d}\right)\times\\
\left[i\sin\left(k_{z}z\right)-\text{sgn}\left(z\right)\left(\frac{1}{v_{1}}-\frac{1}{2}r_{1}k^{2}\right)^{-1}\frac{\sqrt{\pi}\eta^{1/4}\mathcal{R}_{z}}{d}e^{ik_{z}\left|z\right|}\right],\label{eq:3.3}
\end{multline}
where $\mathcal{R}_{z}$ is defined as in Eq.(\ref{eq:2.9}). For
this 1D scattering problem in the waveguide, we anticipate the scattering
wavefunction at large distance takes the form
\begin{equation}
\psi\left(\mathbf{r}\right)\sim\phi_{0}\left(\frac{\sqrt{\eta}x}{d}\right)\phi_{0}\left(\frac{y}{d}\right)\left[i\sin\left(k_{z}z\right)+\text{sgn}\left(z\right)f_{p}e^{ik_{z}\left|z\right|}\right],\label{eq:3.4}
\end{equation}
where $f_{p}$ is the effective $p$-wave 1D scattering amplitude.
Comparing Eqs.(\ref{eq:3.3}) and (\ref{eq:3.4}), we easily obtain
the effective 1D scattering amplitude
\begin{equation}
f_{p}=-\left(\frac{1}{v_{1}}-\frac{1}{2}r_{1}k^{2}\right)^{-1}\frac{\sqrt{\pi}\eta^{1/4}\mathcal{R}_{z}}{d}.\label{eq:3.5}
\end{equation}
The coefficient $\mathcal{R}_{z}$ is determined by the asymptotic
behavior of the scattering wavefunction (\ref{eq:3.1}) at small distance,
i.e., $r\approx0$ , and we find
\begin{multline}
\mathcal{R}_{z}=\frac{i6k_{z}\eta^{1/4}}{\sqrt{\pi}d}\left[1+\frac{6}{\sqrt{\pi}d^{3}}\left(\frac{1}{v_{1}}-\frac{1}{2}r_{1}k^{2}\right)^{-1}\mathcal{F}_{z}^{\prime}\left(\epsilon,0\right)\right.\\
\left.+\frac{i6k_{z}\sqrt{\eta}}{d^{2}}\left(\frac{1}{v_{1}}-\frac{1}{2}r_{1}k^{2}\right)^{-1}\right]^{-1},\label{eq:3.6}
\end{multline}
 where 
\begin{multline}
\mathcal{F}_{z}^{\prime}\left(\epsilon,0\right)=\int_{0}^{\infty}d\tau\left\{ \frac{\sqrt{\eta}\exp\left(\epsilon\tau\right)}{\sqrt{2}\tau^{3/2}}\right.\\
\times\left[\frac{1}{\sqrt{\left(1-e^{-2\eta\tau}\right)\left(1-e^{-2\tau}\right)}}-1\right]-\\
\left.\left[\frac{1}{2\sqrt{2}\tau^{5/2}}+\frac{\epsilon+\left(\eta+1\right)/2-2\sqrt{\eta}}{2\sqrt{2}\tau^{3/2}}\right]\right\} .\label{eq:3.7}
\end{multline}

For the low-energy 1D scattering problem, i.e., $k_{z}d\ll1$ , it
is convenient to use the effective 1D scattering lenght $l_{1D}$
to characterize scattering resonances, which can be defined from the
scattering amplitude at zero energy ($\epsilon\approx0^{+}$) as \cite{Pricoupenko2008R}
\begin{equation}
f_{p}=-\frac{ik_{z}}{ik_{z}+l_{1D}^{-1}},\label{eq:3.8}
\end{equation}
and 
\begin{equation}
\frac{l_{1D}}{d}\equiv6\sqrt{\eta}\left[\frac{d^{3}}{v_{1}}-dr_{1}\cdot\frac{\eta+1}{2}+\frac{6}{\sqrt{\pi}}\mathcal{F}_{z}^{\prime}\left(0,0\right)\right]^{-1}.\label{eq:3.9}
\end{equation}
The divergence of the 1D scattering length $l_{1D}$ characterizes
the scattering resonance in the waveguide. By tuning the 3D scattering
volume $v_{1}$ and the effective range $r_{1}$ according to the
$p$-wave Feshbach resonance, the $p$-wave CIR can be reached in
the 1D waveguide when
\begin{equation}
\frac{d^{3}}{v_{1}}-dr_{1}\cdot\frac{\eta+1}{2}=-\frac{6}{\sqrt{\pi}}\mathcal{F}_{z}^{\prime}\left(0,0\right).\label{eq:3.10}
\end{equation}
 For a transversely isotropic confinement, i.e., $\eta=1$ , the 1D
resonance condition (\ref{eq:3.10}) returns to the well-known result
\cite{Granger2004T,Pricoupenko2008R} 
\begin{equation}
\frac{d^{3}}{v_{1}}-dr_{1}=12\zeta\left(-\frac{1}{2}\right),\label{eq:3.11}
\end{equation}
 where $\zeta\left(\cdot\right)$ is the Riemann Zeta function, and
note that we include the 3D $p$-wave effective range in our expression.

\begin{figure}
\includegraphics[width=1\columnwidth]{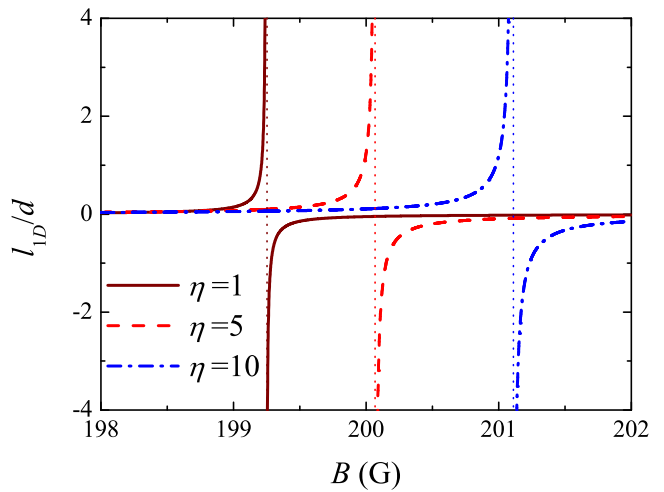}

\caption{(Color online) The $p$-wave 1D scattering length $l_{1D}$ as functions
of the magnetic field $B$ in the 1D waveguide with different transverse
anisotropy $\eta$ . Here, the $p$-wave interaction of two $^{40}$K
atoms in the $\left|F=\frac{9}{2},m_{F}=-\frac{7}{2}\right\rangle $
hyperfine state is tuned by using the $p$-wave Feshbach resonance
centered at $B_{0}=198.8$G in three dimension \cite{Ticknor2004M,Gunter2005P}.}

\label{1DScatteringLength}
\end{figure}

\begin{figure}
\includegraphics[width=1\columnwidth]{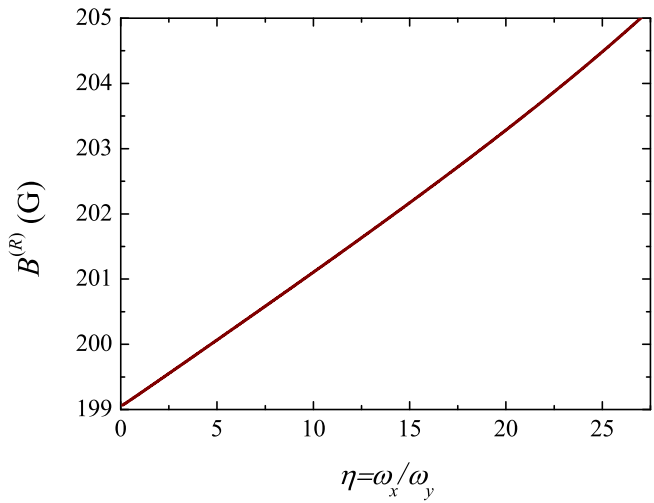}

\caption{(Color online) The position of the $p$-wave confinement-induced resonance
in the 1D waveguide denoted by the magnetic field strength $B^{(R)}$
as a function of the transverse anisotropy $\eta$ . In this plot,
we tune the transversely anisotropy $\eta$ by increasing $\omega_{x}$
with fixed $\omega_{y}$ . The corresponding parameters are still
from the $^{40}$K experiments \cite{Ticknor2004M,Gunter2005P}.}

\label{PositionOfResonance}
\end{figure}

In Fig.\ref{1DScatteringLength}, we present the 1D scattering length
$l_{1D}$ as function of the magnetic field strength $B$ in the waveguide
with three typical transverse anisotropies $\eta=1,\,5,\,\text{and}\,10$.
Here, we still consider two $^{40}$K in the $\left|F=\frac{9}{2},m_{F}=-\frac{7}{2}\right\rangle $
hyperfine state near the $p$-wave Feshbach resonance centered at
$B_{0}=198.8$G in 3D \cite{Ticknor2004M,Gunter2005P}. We find that,
as the transverse anisotropy $\eta$ increases, the resonance position
of $p$-wave CIR shifts to a higher magnetic field strength. More
apparently, we show how the resonance position denoted by the magnetic
field strength $B^{(R)}$ shifts with the transverse anisotropy $\eta$
in Fig.\ref{PositionOfResonance}.

\section{The effective two-channel mechanism\label{sub:TheEffectiveTwoChannelMechanism}}

For $s$-wave CIR in a quais-1D waveguide, there is a simple two-channel
picutre first introduced by Bergeman \emph{el al}. \cite{Bergeman2003A},
and then extended to the case of the transversely anisotropic confinement
\cite{Peng2010C}. We may understand this two-channel mechanism as
follows: due to the low temperature and tight transverse confinement,
the two atoms can only enter from the transverse ground state, while
the transverse excited modes are all closed. Therefore, the transverse
ground state and the manifold of the remaining transverse excited
states may be regarded as the open and closed channels, respectively.
If a molecular state exists in the closed channel, a zero-energy scattering
resonance occurs when this molecule energetically coincides with the
continuum threshold of the open channel. In this section, we aim to
study whether this simple two-channel picture is still valid for $p$-wave
interaction.

Owing to the separability of c.m. motion and relative motion in harmonic
traps, we still only focus on the relative-motion Hamiltonian (\ref{eq:2.1}).
According to the two-channel mechanism proposed for $s$-wave interaction
\cite{Bergeman2003A}, the total relative-motion Hamiltonian $\mathcal{H}$
may formally be splitted into three terms, i.e., $\mathcal{H}_{op}$
, $\mathcal{H}_{cl}$ and $\mathcal{W}$ corresponding to the open
channel, closed channel, and coupling part, respectively,
\begin{eqnarray}
\mathcal{H} & = & \mathcal{H}_{op}+\mathcal{H}_{cl}+\mathcal{W}\nonumber \\
 & \equiv & P_{g}\mathcal{H}P_{g}+P_{e}\mathcal{H}P_{e}+\left(P_{g}\mathcal{H}P_{e}+H.c\right),\label{eq:4.1}
\end{eqnarray}
where $P_{g}=\left|g\right\rangle \left\langle g\right|$ , $P_{e}=\sum_{\alpha\neq g}\left|\alpha\right\rangle \left\langle \alpha\right|$
are the corresponding projection operators, and $\left|g\right\rangle $
, $\left|\alpha\right\rangle $ are the transverse ground and excited
states, respectively. In the follows, we are going to show that the
crossing of the molecular state in the closed channel $\mathcal{H}_{cl}$
and the energy continuum threshold of the open channel, i.e., $\hbar\omega\left(\eta+1\right)/2$
, denotes the position where the $p$-wave CIR occurs.

\begin{figure}
\includegraphics[width=1\columnwidth]{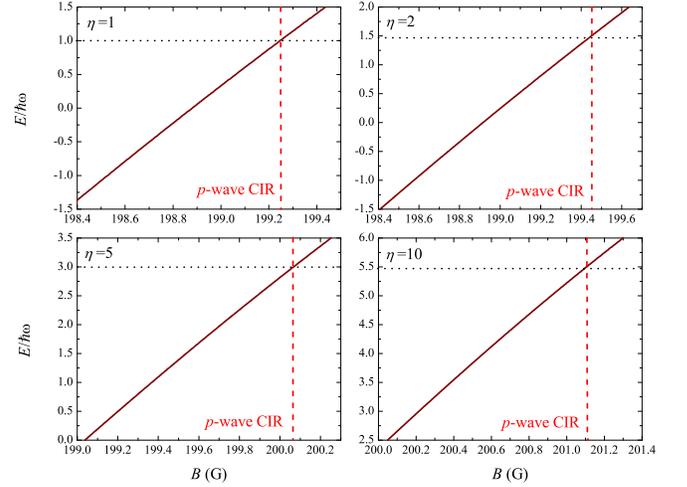}

\caption{(Color online) The molecular state of the closed channel Hamiltonian
$\mathcal{H}_{cl}$ for different transverse anisotropy $\eta$ .
The horizontal dotted lines denote the continuum threshold of the
open channel, while the vertical red dashed lines indicate the position
where a zero-energy scattering resonance occurs.}

\label{ExcitedBoundState}
\end{figure}

In order to obtain the molecular state of the closed channel, we need
to solve the Sch\"{o}rdinger equation $\mathcal{H}_{cl}\psi_{cl}\left(\mathbf{r}\right)=E\psi_{cl}\left(\mathbf{r}\right)$
. The procedure is similar to that of obtaining the bound states of
the full Hamiltonian (\ref{eq:2.1}). However, we need to project
out the transverse ground state, and this means the transverse ground
state, i.e., the $n_{1}=n_{2}=0$ term, should be excluded in the
expansion of the Green's function, i. e., Eq.(\ref{eq:2.6}) . Similarly,
we obtain the Green's function for the closed channel Hamiltonian,\begin{widetext}

\begin{multline}
G_{E}^{(cl)}\left(\mathbf{r},\mathbf{r}^{\prime}\right)=\frac{\sqrt{\eta}}{\pi^{3/2}d^{3}\hbar\omega}\exp\left[-\frac{\eta\left(x^{2}+x^{\prime2}\right)+y^{2}+y^{\prime2}}{2d^{2}}\right]\int_{0}^{\infty}d\tau\frac{e^{\epsilon\tau}}{\sqrt{2\tau}}\exp\left[-\frac{\left(z-z^{\prime}\right)^{2}}{2d^{2}\tau}\right]\\
\times\left\{ \frac{\exp\left[\eta\frac{2xx^{\prime}-\left(x^{\prime2}+x^{2}\right)e^{-\eta\tau}}{2d^{2}\sinh\left(\eta\tau\right)}+\frac{2yy^{\prime}-\left(y^{\prime2}+y^{2}\right)e^{-\tau}}{2d^{2}\sinh\tau}\right]}{\sqrt{\left(1-e^{-2\eta\tau}\right)\left(1-e^{-2\tau}\right)}}-1\right\} .\label{eq:4.2}
\end{multline}
\end{widetext} Then the bound-state wavefunction $\psi_{cl}\left(\mathbf{r}\right)$
takes the form
\begin{multline}
\psi_{cl}\left(\mathbf{r}\right)=-\frac{\pi\hbar^{2}}{\mu}\left(\frac{1}{v_{1}}-\frac{1}{2}r_{1}k^{2}\right)^{-1}\\
\cdot\sum_{n}\mathcal{R}_{n}^{\prime}\left[\frac{\partial}{\partial r_{n}^{\prime}}G_{E}^{(cl)}\left(\mathbf{r},\mathbf{r}^{\prime}\right)\right]_{r^{\prime}=0},\label{eq:4.3}
\end{multline}
 where
\begin{equation}
\mathcal{R}_{n}^{\prime}\equiv\left[\frac{\partial^{3}}{\partial r^{3}}r^{3}\frac{\partial}{\partial r_{n}}\psi_{cl}\left(\mathbf{r}\right)\right].\label{eq:4.4}
\end{equation}
 After straightforward and similar algebra as that in Sec.\ref{sec:BoundStateSolutions},
we find there are also three bound states in the closed channel, and
two of them are with odd transverse parity, while that of the third
is even, which takes the form,
\begin{equation}
\psi_{cl,z}\left(\mathbf{r}\right)=z\left[\frac{1}{r^{3}}+\frac{E}{\hbar\omega}\frac{1}{rd^{2}}+\frac{2}{\sqrt{\pi}d^{3}}\mathcal{F}_{z}^{\prime}\left(\epsilon,\mathbf{r}\right)\right],\label{eq:4.5}
\end{equation}
where\begin{widetext}
\begin{multline}
\mathcal{F}_{z}^{\prime}\left(\epsilon,\mathbf{r}\right)=\int_{0}^{\infty}d\tau\left\{ \frac{\sqrt{\eta}\exp\left(\epsilon\tau-\frac{z^{2}}{2d^{2}\tau}-\frac{\eta x^{2}+y^{2}}{2d^{2}}\right)}{\sqrt{2}\tau^{3/2}}\left[\frac{\exp\left(-\frac{\eta x^{2}e^{-\eta\tau}}{2d^{2}\sinh\left(\eta\tau\right)}-\frac{y^{2}e^{-\tau}}{2d^{2}\sinh\tau}\right)}{\sqrt{\left(1-e^{-2\eta\tau}\right)\left(1-e^{-2\tau}\right)}}-1\right]-\right.\\
\left.\frac{1}{2\sqrt{2}}\left[\frac{1}{\tau^{5/2}}+\left(\frac{E}{\hbar\omega}-2\sqrt{\eta}\right)\frac{1}{\tau^{3/2}}\right]e^{-\frac{r^{2}}{2d^{2}\tau}}\right\} .\label{eq:4.6}
\end{multline}
\end{widetext} The corresponding binding energy of the bound state
with even transverse parity satisfies,
\begin{equation}
\frac{d^{3}}{v_{1}}=-\frac{6}{\sqrt{\pi}}\mathcal{F}_{z}^{\prime}\left(\epsilon,0\right)+dr_{1}\cdot\frac{E}{\hbar\omega}.\label{eq:4.7}
\end{equation}
Obviously, $\mathcal{F}_{z}^{\prime}\left(\epsilon,0\right)$ takes
the form of Eq.(\ref{eq:3.7}). If two atoms enter from the transverse
ground state (open channel) with energy below the transverse excited
states (closed channel), they may only be coupled to the molecular
state $\psi_{cl,z}\left(\mathbf{r}\right)$ with even transverse parity,
due to the parity conservation. When this molecular state energetically
crosses the continuum threshold of the open channel ($\epsilon\approx0$),
a zero-energy scattering resonance occurs. From Eq.(\ref{eq:4.5}),
the resonance condition is given by
\begin{equation}
\frac{d^{3}}{v_{1}}-dr_{1}\cdot\frac{\eta+1}{2}=-\frac{6}{\sqrt{\pi}}\mathcal{F}_{z}^{\prime}\left(0,0\right),\label{eq:4.8}
\end{equation}
 which returns to Eq.(\ref{eq:3.10}) again. The binding energy of
the $^{40}$K-$^{40}$K molecular state $\psi_{cl,z}\left(\mathbf{r}\right)$
in the closed channel $\mathcal{H}_{cl}$ under different transverse
anisotropy is illustrated in Fig.\ref{ExcitedBoundState}, and the
interatomic interaction is still tuned by $p$-wave Feshbach resonance
centered at $B_{0}=198.8$G. The vertical dashed lines indicate where
the $p$-wave CIR occurs.

\section{Conclusions\label{sec:Conclusions}}

We theoretically investigate the influence of the transverse anisotropy
of the confinement on the two-body state with $p$-wave interaction
in a quasi-one-dimensional waveguide. The two-body problem in such
quasi-one-dimensional systems is solved. The interatomic interaction
is modeled by $p$-wave pseudopotential. We find there are totally
three bound states due to the non-zero orbital angular momentum of
$p$-wave interaction, while there is only one two-body bound state
supported by $s$-wave pseudopotential. In addition, the effective
one-dimensional scattering amplitude and scattering length are derived.
We predict the $p$-wave confinement-induced resonance for any transverse
anisotropy of the waveguide, whose position shows an apparent shift
with increasing transverse anisotropy. Besides, a two-channel mechanism
is presented for $p$-wave confinement-induced resonance in a waveguide,
which was first proposed for $s$-wave interaction. We find this effective
two-channel picture is still valid for $p$-wave interaction. All
our calculations are based on the parameterization of the $^{40}$K
experiments, and can be confirmed in future experiements.
\begin{acknowledgments}
This work is supported by NSFC (Grants No. 11434015, No. 11204355,
No. 11474315 and No. 91336106), NBRP-China (Grant No. 2011CB921601),
CPSF (Grants No. No. 2013T60762) and programs in Hubei province (Grants
No. 2013010501010124 and No. 2013CFA056).\end{acknowledgments}

\end{document}